\documentclass[10pt,letterpaper]{article}
\usepackage[top=0.85in,left=2.75in,footskip=0.75in,marginparwidth=2in]{geometry}

\usepackage[utf8]{inputenc}

\usepackage{cite}

\usepackage{nameref,hyperref}

\usepackage[right]{lineno}

\usepackage{microtype}
\DisableLigatures[f]{encoding = *, family = * }

\raggedright
\usepackage{changepage}

\usepackage{amsmath}

\usepackage[aboveskip=1pt,labelfont=bf,labelsep=period,singlelinecheck=off]{caption}

\makeatletter
\renewcommand{\@biblabel}[1]{\quad#1.}
\makeatother

\usepackage{lastpage,fancyhdr,graphicx}
\usepackage{epstopdf}
\fancyheadoffset[L]{2.25in}
\fancyfootoffset[L]{2.25in}

\usepackage{color}

\definecolor{Gray}{gray}{.25}

\usepackage{graphicx}

\usepackage{sidecap}

\usepackage{wrapfig}
\usepackage[pscoord]{eso-pic}
\usepackage[fulladjust]{marginnote}
\reversemarginpar

\begin{document}
\vspace*{0.35in}

\begin{flushleft}
{\Large
\textbf\newline{Physics-Informed Optimisation of Conveyor Mode Spin Qubit Transport}
}
\newline
\\
Andrii Sokolov\textsuperscript{1,*},
Conor Power\textsuperscript{1,2},
Elena Blokhina\textsuperscript{1,2},
\\
\bigskip
\bf{1} Equal1 Laboratories, Dublin, D04 V2N9, Ireland
\\
\bf{2} School of Electrical and Electronic Engineering, University College Dublin, Dublin, D04 V1W8, Ireland
\\
\bigskip
* andrii.sokolov@equal1.com

\end{flushleft}

\section*{Abstract}
Scalable quantum information processing in spin-based architectures necessitates the ability to reliably shuttle quantum states across extended device regions with minimal decoherence. In this work, we present a physics-informed algorithm for optimizing electrostatic bias sequences that enable conveyor-mode electron transport in silicon-based quantum dot devices. Our approach combines self-consistent Poisson and Schr\"odinger solvers to maintain a constant ground state energy and enable near-constant velocity shuttling, with potential applicability to both single-electron and hole transport. We validate the algorithm across three representative technologies: Fully-Depleted Silicon on Insulator (FD-SOI), Silicon Metal-Oxide-Seminconductor (SiMOS) and Silicon-Germanium Heterostracture (Si/SiGe), highlighting key limitations and material-specific effects that influence transport fidelity. Our findings underscore the impact of gate geometry, dielectric interfaces, and quantum dot size on the stability of shuttling operations, and offer pathways toward improving coherence preservation in large-scale quantum systems.


\section{\label{sec:introduction} Introduction}

Shuttling operations are fundamental to the functionality of future spin-based quantum processors~\cite{Kunne2024_SpinBus},~\cite{Langrock2023_BlueprintShuttle},~\cite{Patomaki2024_PipelineSpin}. Due to practical limitations in the spatial density of sensors, vias and control gates, direct dot-to-dot coupling across extended quantum devices is not practical and feasible for large scale architectures. Instead, quantum states must be transferred between distant sites, necessitating reliable and coherent shuttling mechanisms~\cite{DeSmet2025_HighFidelityShuttling},~\cite{Seidler2022_ConveyorSiSiGe},~\cite{Nemeth2024_OmnidirectionalShuttling}. Two critical parameters define the performance of any shuttling system: the shuttling duration, which should be significantly shorter than the qubit coherence time, and the degree of additional decoherence introduced during transfer~\cite{Foster2025_DephasingShuttle}.

Two primary modes of shuttling have been established in the literature: the \emph{bucket brigade} approach~\cite{Mills2019_BucketBrigade},~\cite{Zwerver2023_SpinShuttling}, in which quantum states are sequentially tunnelled between adjacent quantum dots, and the \emph{conveyor mode}~\cite{Seidler2022_ConveyorMode},~\cite{Xue2024_QubusConveyor}, wherein a quantum dot or cavity is moved electrostatically across the device without tunnelling. Among these, the conveyor mode has demonstrated lower decoherence levels~\cite{DeSmet2025_HighFidelityShuttling}, making it a more promising candidate for scalable implementations.

Experimentally, the conveyor mode is typically implemented by applying phase-shifted, periodic voltages to a series of control terminals~\cite{DeSmet2025_HighFidelityShuttling},~\cite{Jeon2025_ArchitecturesPulses}. This generates a moving potential that carries the quantum state across the system. While sine waveforms are commonly used due to their simplicity, they may not represent the most optimal control strategy for minimising decoherence and ensuring fidelity~\cite{Nagai2025_DigitalControlledConveyor}.

 The shape of the optimal signal will strongly depend on the geometry due to the gate deposition process. Accurate modelling of quantum state shuttling represents a significant theoretical and computational challenge. Unlike stationary qubits, a shuttled electron experiences perturbations due to time-dependent potential well motion in which both orbital and spin degrees of freedom evolve dynamically. The resulting dynamics are governed by a Hamiltonian that depends explicitly on position and time, making it difficult to apply standard static or adiabatic approximations. Earlier demonstrations of charge and spin transfer through multi-dot arrays have already highlighted the need for time-dependent models that can capture both tunnelling and motional coherence effects~\cite{Mills2019_BucketBrigade},~\cite{Zwerver2023_SpinShuttling}.

One major source of computational complexity is due to the multi-scale nature of the problem. The shuttling process involves nanosecond-scale gate control dynamics with the orbital, spin precession and spin relaxation processes. Capturing these simultaneously in simulations requires extremely fine time steps for temporal resolution. This leads to extremely demanding numerical requirements. Furthermore realistic simulations should include more effects, such as  electrostatic cross-talk, charge noise and valley–orbit coupling which modify the effective potential in real time as the electron moves between gates~\cite{Ermoneit2024_OptimalControlShuttling},~\cite{Seidler2022_ConveyorMode}.

Connecting models to experiment is hindered by limited knowledge of device-specific parameters such as dielectric constants, gate capacitances, and background charge traps. As a result, calibration-based semi-empirical models are typically used, but these obscure the underlying physical mechanisms. Achieving predictive power therefore requires coupling self-consistent electrostatic simulations with time-dependent quantum dynamics, an approach that remains computationally expensive even for few-dot systems~\cite{Nagai2025_DigitalControlledConveyor}. 

This work addresses some of the major challenges of optimising control signals for conveyor mode shuttling by integrating techniques from semiconductor physics and quantum mechanics. In particular, we want to address large-scale simulations for shuttling through a long conveyor bus where gates may be deposited non-evenly.  The proposed simulation methodology will not assume a fixed shape of a potential well and will be resolved through self-consistent simulations. The methodology can be readily applied to various geometries and processes. In particular, we use on the most common spin qubit platform~\cite{GonzalezZalba2021_ScalingCMOS},~\cite{Kawakami2016_GateFidelity},~\cite{Huang2019_TwoQubitFidelity},~\cite{Maurand2016_CMOSSpinQubit},~\cite{power2025fully}: SiGe heterostructure, Silicon Metal-Oxide Semiconductor (SiMOS) and Fully-Deplted Silicon-on-Insulator FDSOI.  In Section~\ref{sec:algorithm}, we present an algorithm designed to generate optimal waveforms for electrostatic quantum transport. Section~\ref{sec:examples} provides examples of the algorithm applied to various device architectures. 

\section{Algorithm\label{sec:algorithm}}
\subsection{Modelling and Simulation Framework}

In this work, we rely in part on the QTCAD$^{\text{\textregistered}}$ simulation toolkit \cite{beaudoin2022robust, prentki2023robust, philippopoulos2024analysis} to carry out a comprehensive set of quantum device modelling tasks. This toolkit enables the self-consistent solution of the Poisson and Schr\"odinger equations to determine the electrostatic potential landscape and quantum confinement properties within semiconductor nanostructures. QTCAD$^{\text{\textregistered}}$ provides a finite-element-based framework capable of capturing realistic device geometries, material interfaces, and doping profiles, allowing accurate modelling of charge distribution, quantum dot formation, and tunnel coupling. In our workflow, it is particularly used to extract key quantities such as energy spectra, wavefunctions, and electrochemical potentials under applied bias and gate voltages, which serve as inputs for subsequent quantum transport and qubit-level simulations.

\begin{figure}[thb]
\centering
\includegraphics[width=0.95\linewidth]{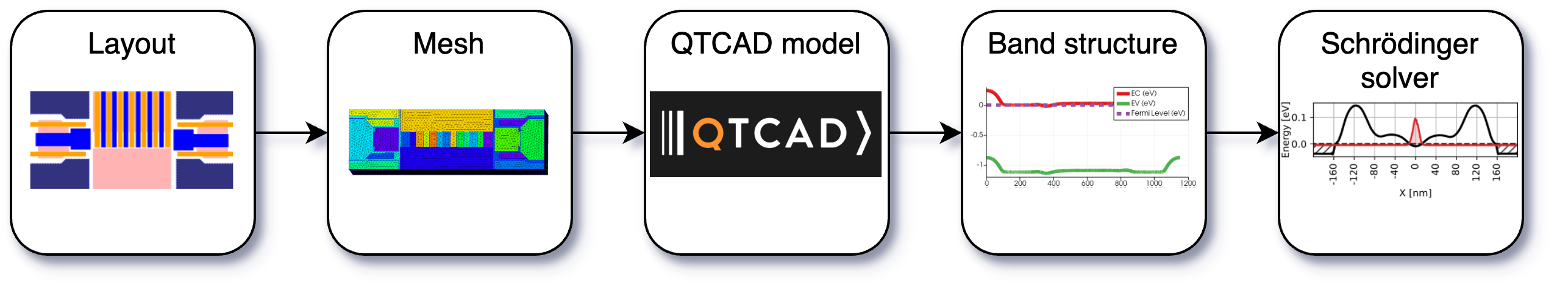}
\caption{The symbolic workflow of the used framework.\label{fig:framework}}
\end{figure}

Specifically, our workflow includes the following operations (Fig.\,\ref{fig:framework}): 
\begin{enumerate}
\item Begin with the 2D layout of the quantum processor 
\item Build 3D model of the given structure with the process layer-stack and TEM of fabricated structures 
\item Create a QTCAD$^{\text{\textregistered}}$ application that includes all properties of the used materials and boundaries.
\item Solve the Poisson equation under cryogenic conditions \cite{ashcroft_mermin_1976}. 
\begin{equation}\label{eqn:poisson}
-\nabla\cdot \left( \varepsilon\nabla\varphi \right)=e\left(p-n + N_+ - N_-\right) + \rho_0,
\end{equation}
where $\varepsilon$ is the dielectric permittivity of the domain, $\varphi$ is the electric potential, $e$ is the elementary charge ($e>0$), $p$ and $n$ are the hole and electron densities, respectively, and $N_+$ and $N_-$ are the densities of ionized donors and acceptors respectively. 
\item Solve the effective stationary Schr\"odinger equation \cite{winkler_2003}, including extraction of eigenenergies, eigenfunctions and properties of quantum dots relevant to electron shuttling operations:
\begin{equation}\label{eqn:schrodinger}
V_\text{conf}(\vec{r})\psi(\vec{r}) - \frac{\hbar^2}{2}\nabla\cdot \mathbf{M}_e^{-1}\cdot\nabla\psi(\vec{r}) = E\psi(\vec{r}),
\end{equation}
where $\psi(\vec{r})$ is the envelope eigenfunction,  $V_\text{conf}(\vec{r})$ is the total confinement potential, $\hbar$ is the reduced Planck constant, $\mathbf{M}_e^{-1}$ is the electron inverse effective mass tensor.
\end{enumerate}

This workflow requires a novel integration of multi-domain simulation tools adapted to the specific requirements of semiconductor spin-qubit design with a very-large computational volume. While some individual physical solvers are known and established, their continuous workflow combined use in a single, self-consistent framework represents a significant methodological advance. By connecting process-based 3D reconstruction, electrostatic modelling and quantum-mechanical analysis within QTCAD$^{\text{\textregistered}}$, we bridge the gap between device fabrication data and qubit-level performance metrics. This multi-domain adaptation enables predictive modelling of charge localisation and transfer dynamics that is directly comparable to low-temperature experimental measurements.

Given the computationally intensive nature of these simulations, especially when high-resolution spatial discretisation is involved, it is crucial to optimise the computational mesh to minimise processing time without compromising accuracy.

\subsection{Bias Search Algorithm}

\begin{figure}[thb]
\centering
\includegraphics[width=0.95\linewidth]{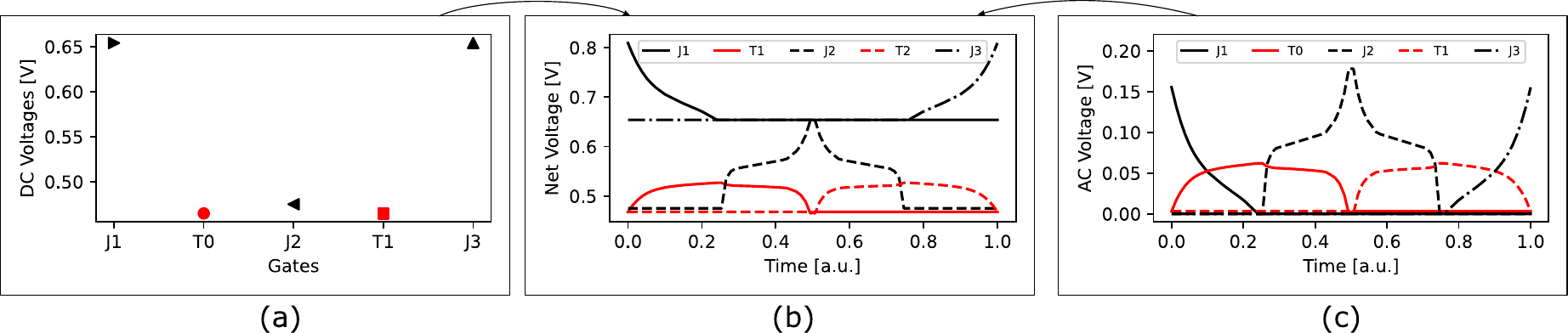}
\caption{\label{fig:acdc} AC and DC voltages applied to the shuttling gates. (a) A DC voltage that provides a flat conduction band. (b) Net AC and DC voltages that create a conveyor mode shuttling. (c) An AC voltage is applied to the gates. }
\end{figure}

The simulation framework from the previous section has been used to implement a novel algorithm for the bias search and shuttling pulses optimisation. The procedure to identify optimal AC and DC biasing (Fig.\,\ref{fig:acdc}) conditions for the conveyor mode shuttling is as follows: 

\begin{enumerate}
\item Search for the optimal DC voltage (flat-band).
\item Identify the individual voltages that one needs to apply to each gate to form the quantum dot below it with a given ground state energy.
\item For each odd gate, form the linearly distributed bias voltages (from maximal to minimal) and compute such voltages on the next gate to keep the ground state energy of the obtained quantum dot constant. \label{algo:unit}
\item As a result of the previous steps, there should be a table of bias voltages for each discrete moment for the conveyor mode shuttling (if it is possible for the given device). As the next step, one needs to run a Poisson and Schr\"odinger solver for each discrete time moment and analyse the energy gaps between the ground state and the first excited state, the dot size and the highest probability to observe the particle in the dot. \label{algo:dataproc} 
\item Since in the Step~\ref{algo:unit}, as a linear function for the bias voltages was used, the shuttling speed for the conveyor is not constant. To make it constant, one needs to build the $V(x)$ function from the Step~\ref{algo:dataproc} data, and uniformly discretise it. This will produce a constant velocity, but does not necessarily keep the ground state energy constant. After the functions for the odd gates are obtained, the voltages for even gates are then calculated similarly to Step~\ref{algo:unit}.
\end{enumerate}

\subsection{DC bias search}\label{sec:optim_DC}

The main goal of the DC bias calculation is to find a set of voltages that creates a flat conduction band edge spaced at a given value from the Fermi level. The DC bias should be great enough to reduce the AC amplitudes applied to the gates, and therefore, reduce the device heating during the shuttling. On the other hand, it should be low enough to create well-formed quantum dots under the gates and between the gates, with the application of AC voltages. 

\begin{figure}[thb]
\centering
\includegraphics[width=0.95\linewidth]{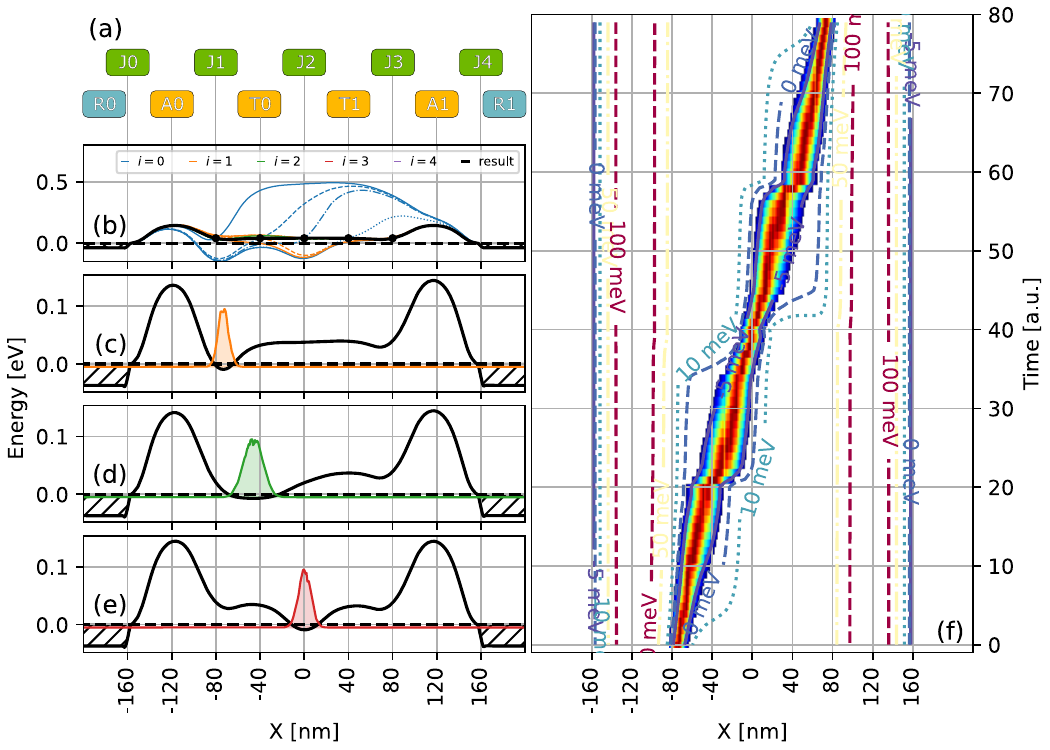}
\caption{\label{fig:algo} An example that shows how different stages of the algorithm works. (a) The cartoon layout of the test structure. R0 and R1 represent the reservoirs; A0 and A1 are the accumulation gates; T0 and T1 are the tunnelling gates; J0, J1, J2, J3 and J4 are the J-gates. (b) Different stages of the DC bias definition algorithm. Different colours represent different iterations of the algorithm from 0 to 4. The thick black line shows the result. Black circles on the graph represent the points selected for the optimisation. (c) The quantum dot formed under the J1 gate (the wavefunction is normalised to be visible). (d)  The quantum dot formed under the T0 gate (the wavefunction is normalised to be visible). (e) The quantum dot formed under the J2 gate (the wavefunction is normalised to be visible). (f) The time-evolution of both conduction band edge --- the contour plot, and the probability density function --- the colour mesh. }
\end{figure}

In this research, we employed two different methods to calculate the flat-band DC voltages. At first, the lever-arm approach was tested. This method was based on the assumption that the energy of the quantum dot linearly depends on the voltages applied to the gate that is just above it, and the neighbouring gates: 
\begin{equation}
E_i = \alpha^i_{(i-1)} V_{i-1} + \alpha^i_i V_i + \alpha^i_{(i+1)}V_{(i+1)} \label{eqn:lever_arm}
\end{equation}
It is possible, then, to build a system of $N$ equations \eqref{eqn:lever_arm} with $N$ unknown voltages that potentially make the energy of quantum dots under each gate equal. This approach has very low computational complexity. However, realistically, this approach is not stable as the energies of the quantum dots are very close, and it is challenging to control the position of quantum dots during the initial and iterative lever arm calculation. 

Thus, the method of simple iterations was used for this problem. In this method, there is a geometrical point selected in the middle of each quantum dot, and the conduction $E_{Ci}$ (or valence $E_{Vi}$) band edge is computed at these points (see black circles in Fig.~\ref{fig:algo}\,b). Then, for each gate, the voltage is optimised in such a way that the corresponding conduction band edge is equal to the target value $E_{C}$. The most efficient way to do this is to create the function:
\begin{equation}
f_i(V_i) \rightarrow E_{Ci}
\end{equation}
Practically, this function includes the numerical finite element method (FEM) solution of the cryogenic semiconductor equation with given voltages, and calculating the $E_{C}$ ($E_{V}$) at the selected point. And numerically solve the equation:
\begin{equation}\label{eqn:ec_equation}
f_i(V_i) - E_{C} = 0
\end{equation}
Since functions $f_i$ are quasi-linear, the most computationally efficient method is the Newton method (see the blue J1~0 line in Fig.~\ref{fig:algo}). It is easy to see that optimising each next gate voltage $V_i$ shifts the previous points on the conduction (valence) band edge curve ($E_{C(i-1)}$, $E_{C(i-2)}$). This happens due to the cross-talk between different gates. However, the cross-talk influence is always lower than the effect of the gate above the quantum dot; therefore, simply iterating the gates, the conduction (valence) band edge will converge to the target value, as shown in Fig.~\ref{fig:algo}\,b. One can spot small parasitic wells on the edge of the conduction band edge that are lower than the target $E_C$. This happens because barriers between the shuttling region and the reservoirs are relatively high, and the conduction band edge has a continuous derivative. The depth of these wells, in fact, dictates the minimal value for the target $E_C$. The good practice is to choose $E_C$ at least twice as high as the parasitic wells forming. Otherwise, the parasitic quantum wells forming there will force the particle to tunnel. 

\subsection{Optimised AC bias search}\label{sec:optim_AC}

After the DC bias voltages are calculated, they can be considered as minimal voltages applied to gates. The AC voltages applied on top should ideally form a single quantum well (or chain of quantum wells) and move it in the shuttling direction with minimal possible distortions. The goal of the Section~\ref{sec:optim_AC} is to describe the algorithm to compute the amplitude and the waveform of such an AC signal. 

Initially, it is necessary to calculate the amplitude of the AC voltage. To do this, the target ground-state energy $E_0^\mathrm{T}$ of the quantum dot should be selected. The good practice is to select it as
\begin{equation}
E_0^\mathrm{T} = -0.5\cdot\left( E_1 - E_0 \right)
\end{equation}
for the formed quantum dot. 

The ground state energy of the quantum dot can be calculated by solving the Poisson equation for the applied voltages, and then solving the Schr\"odinger equation for the obtained potential energy. This can be written as the function:
\begin{equation}
g_i(V_i) \rightarrow E_0^i
\end{equation}
For every gate, the maximal voltage can be defined by solving the equation:
\begin{equation}\label{eqn:eo_equation}
g_i(V_i) - E_0^\mathrm{T} = 0
\end{equation}
As in the case of the equation \eqref{eqn:ec_equation}, this equation is efficiently solved by the Newton method. In Figure~\ref {fig:algo}\,c, an example of the described above algorithm is shown.

The proposed simulation is based on solving the stationary Poisson equation and time-independent Schr\"odinger equation. Therefore, we assume that the shuttling speed is slow enough, and the transition from one state to another happens quasi-adiabatically. In this assumption, the speed of the shuttling is controlled by the time between two states $\Delta t$, which can be scaled. Thus, in all the next graphs, we put the time in arbitrary units (a.u.). 

\begin{figure}[thb]
\centering
\includegraphics[width=0.99\linewidth]{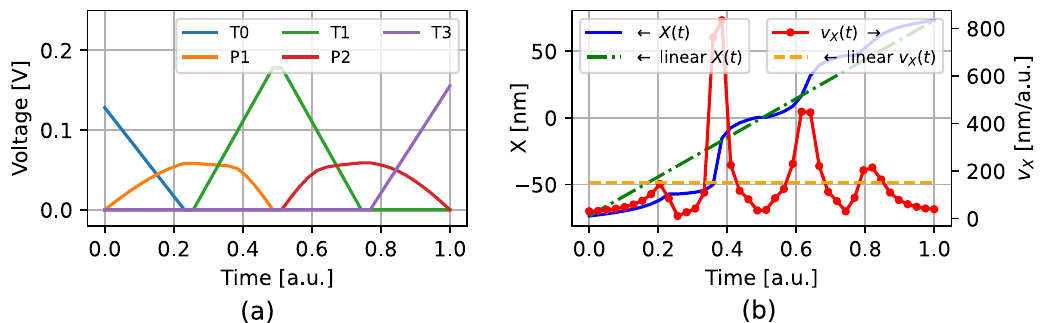}
\caption{\label{fig:algo_speed} An example of the intermediate step of the AC-voltage definition algorithm. (a) Voltages applied to different gates. (b) Coordinate $X(t)$ of the quantum dot --- solid blue line. If the quantum dot were to move with a constant speed, the coordinate would change as shown by the green dash-dot line. The solid red line with circles represents the speed of the quantum dot. The constant speed is the orange dashed line. }
\end{figure}

As it was written in Sec.~\ref{sec:algorithm}, the next step after the computation of the AC voltages is to sample the odd gate voltages from the maximum to the minimum value using the linear function (Fig.~\ref{fig:algo_speed}\,a). For each value of the odd gate voltages, it is possible to solve the equation \eqref{eqn:eo_equation} and find the corresponding voltage for the next even gate. 

With this approach, the shuttling happens in dashes (Fig.~\ref{fig:algo_speed}\,b). In general, in the quasi-adiabatic approach used here for shuttling, there is no difference if the velocity of the quantum dot is not constant. However, if one wants to extend this approach to higher speeds, it is better to have a constant shuttling speed as desired by the user. This happens because the voltages on the odd gates were chosen to be linear functions, but to obtain the constant velocity AC biasing, these functions should be set up in a specific way. 

The waveform for the odd numbers of the gates can be calculated from the results presented in Figure~\ref{fig:algo_speed}b. To do this, one needs to define the functions $V_i(X)$ and uniformly discretise these functions. Examples of this can be found in Fig.~\ref{fig:example_fdsoi}b,f and Fig.~\ref{fig:example_simos}b,f. They will be discussed in more detail in the corresponding sections. 

After the odd functions $V_i(t)$, that provide a constant velocity, are calculated for each time step, the even voltages can be calculated by solving equation \eqref{eqn:eo_equation}. 

Applying the steps described above, one can calculate the DC and AC biases that produce conveyor mode shuttling with a constant ground state energy $E_0$ equal to the target value $E_0^\mathrm{T}$ and at a constant velocity $v_x$ if this is possible for the given geometry (Fig.~\ref{fig:algo}\,f). There is a possibility described below, where the shuttling includes tunnelling events; however, the proposed method provides the best chance for the conveyor mode shuttling. 

In addition, above, we described the way to compute a single pulse applied to every gate to form a single well that is moving uniformly. In reality, the pulses should be periodic to create a sequence of quantum wells shuttling with a constant speed. However, assuming that the cross-talk between gate number $i$ and  gate number $i-2$ is negligible, the waveforms should be kept the same as for a single moving well. 

\section{Application of the Methodology to Different Spin Qubit Platforms\label{sec:examples}}

In this work, we used a generic FD-SOI set of transistors and a SiMOS device to test this algorithm. On the basis of the obtained results, we made a conclusion about what influence of the conveyor mode on shuttling stability. 

\subsection{FD-SOI device\label{sec:fdsoi}}

\begin{figure}[thb]
\includegraphics[width=0.99\linewidth]{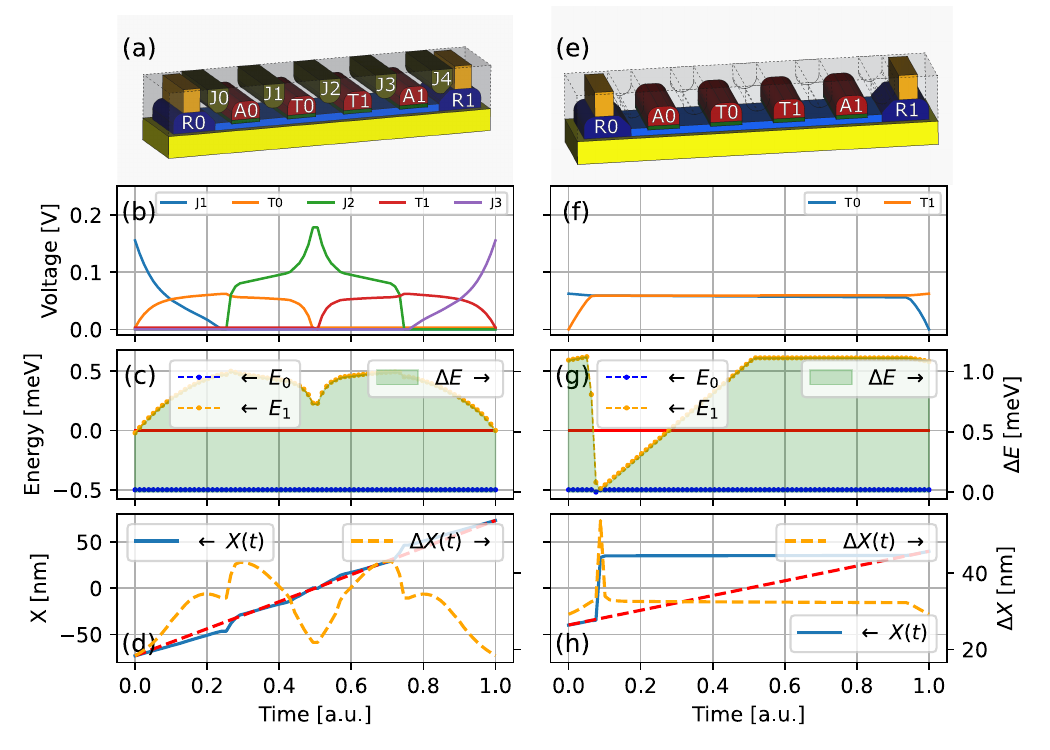}
\caption{\label{fig:example_fdsoi} Summary of the FD-SOI simulations. (a) 3D image of the quantum device: buried oxide --- yellow; raised source and drain that are doped --- deep blue; undoped silicon channel --- light blue; gate oxide --- green; polysilicon gate --- red; tungsten J-gates --- olive; metal contacts --- orange; spacer --- transparent gray. (b) Optimal AC voltages are applied for the different gates for the conveyor mode shuttling. (c) Ground-state energy --- blue dashed line; The first excited state energy --- orange dashed line. Energy difference --- green filling. (d) The coordinate $X$ over time --- blue line, with the constant speed coordinate over time --- red dashed line. The size of the quantum dot $\Delta X$ ---orange dashed line. (e) The version of the same device without J-gates. Here is the same colour legend as in Figure~\ref{fig:example_fdsoi}\,a. (f) Optimised gate voltages applied to the device without J-gates. (g) The ground state energy and the first excited state energy evolution with time. (h) The coordinate and the size of the quantum dot evolution. }
\end{figure}

Fully-depleted silicon-on-insulator (FD-SOI) technologies are gaining attention for quantum applications. However, realistic FD-SOI devices typically exhibit relatively large gate pitches, in the range of 80–100 nm. This substantial spacing between adjacent gates presents a significant challenge for implementing conveyor-mode electron shuttling, as it precludes the proximity required for seamless quantum dot coupling and coherent transport. 

To address this limitation, some research groups have employed post-processing techniques to introduce tungsten J-gates as additional plunger gates on top of pre-fabricated FD-SOI transistor arrays \cite{Bedecarrats_2021}. These plungers act as control gates to modulate the potential landscape along the transport channel. In our study, we instead explore an idealised architecture where the gate pitch is chosen to be 80 nm and tungsten gates are relatively close to the structure, Fig.~\ref{fig:example_fdsoi}a.

In the absence of applied voltages to the gates and plungers, the electrostatic landscape of the device reveals a global potential barrier between the source and drain. This barrier is inherently non-uniform, arising from the varied dielectric properties and work functions of the gate oxide and surrounding spacer materials. Such non-uniformity necessitates the application of the DC voltage alignment to initialise the device into a suitable transport configuration as was described in Section~\ref{sec:optim_DC}.

Figure~\ref{fig:example_fdsoi}\,b shows the optimised AC gate voltage profile derived from our modelling, which enables a quasi-constant electron velocity and maintains a nearly constant ground state energy during the shuttling process. In Figure~\ref{fig:example_fdsoi}\,c, it is easy to see that for the voltages obtained by the algorithm, the energy spacing between the ground state and the first excited state is higher than 0.5\,meV for the shuttling time. The speed during shuttling is also constant within the numerical error (Fig.~\ref{fig:example_fdsoi}\,d).

However, it is not always possible to achieve the conveyor mode shuttling even if the algorithm is implemented successfully. To demonstrate this, the same device was used, but without J-gates (Fig.~\ref{fig:example_fdsoi}\,e). Figure~\ref{fig:example_fdsoi}\,f shows the voltages applied to the T0 and T1 gates. The AC voltage is lower than in the previous case, because the voltage applied to the accumulation gates was chosen slightly higher. 

The analysis of the energy gap shows that the energy gap becomes zero at approximately 0.1 a.u. of time (see Fig.~\ref{fig:example_fdsoi}\,g). Physically, this means that there is a tunnelling event at this time point that is visible from Fig.~\ref{fig:example_fdsoi}\,h. The coordinate $X(t)$ looks like a step-function at this point, and the size of the dot $\Delta X(t)$ has a spike at the same point. Also, it is easy to see that at the beginning and the end of the process, the electron is moving with a constant speed; however, after a certain point, there is a tunnelling that looks like a step-function. 

\subsection{SiMOS device\label{sec:simos}}

\begin{figure}[thb] 
\includegraphics[width=0.95\linewidth]{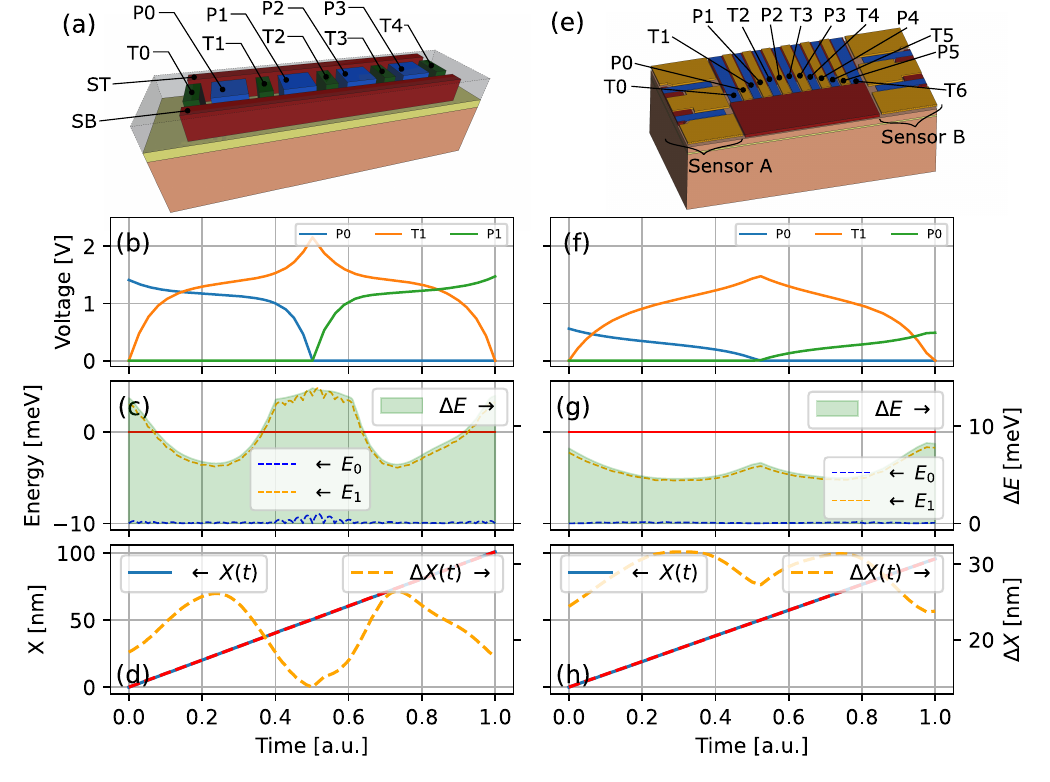} \caption{\label{fig:example_simos} The summary of the SiMOS and SiGe simulations. (a) The layout for the SiMOS device without sensors. Bulk silicon wafer --- peach colour; silicon oxide layer --- yellow; screening gates --- red; plunger gates --- blue; tunnelling gates --- green. (b) Optimal AC waveforms to shuttle from the gate P0 to the gate P1 for SiMOS device calculated applying the algorithm from Section~\ref{sec:algorithm}. (c) The ground-state energy --- blue dashed line, the energy of the first excited state --- orange dashed line, zero energy --- red solid line, along with the energy gap --- green filling over time. (d) The change of the most-probable coordinate of an electron $X(t)$ --- blue dashed line and constant-speed line --- red line over time. The change of the quantum dot size $\Delta X(t)$ over time --- orange dashed line. (e) The 3D model of the Si/SiGe device. The silicon layer --- yellow; the SiGe layer --- peach. Gates are deposited in three different layers: the first layer --- red; the second layer --- orange, and the third layer --- blue. (f) Optimal AC waveforms for Si/SiGe device. (g) The energy diagram with a colour scheme like in (b). (h) The shuttling coordinates with colours like in (d). } 
\end{figure}

Recent work has demonstrated the feasibility of fabricating regular gate structures in SiMOS devices using a two-level VIA architecture \cite{PhysRevB.111.195302}. These structures are particularly well-suited for electron shuttling applications. To evaluate the shuttling algorithm described above, we implemented a toy model based on such a SiMOS device. The model consists of 50 nm × 50 nm plunger gates and 20 nm × 50 nm tunnelling gates, as illustrated in Fig.~\ref{fig:example_simos}a.

Due to the relatively small size of the quantum dots formed in this structure, the energy level spacing is larger than in the FD-SOI implementations described in Section~\ref{sec:fdsoi}. Accordingly, the ground state energy was set to $-10$ meV for the simulations. Using the proposed algorithm, the bias voltages were optimised to achieve constant-velocity electron shuttling while maintaining a nearly constant ground state energy, as shown in Fig.~\ref{fig:example_simos}b.

The resulting voltage waveforms for both the tunnelling and plunger gates exhibit similar shapes. However, the amplitude of the tunnelling gate voltage modulation is slightly lower than that of the plunger gates. This difference arises because the plunger gates are geometrically larger than the tunnelling gates, leading to stronger coupling to the quantum dot.

The spatial evolution of the electron's ground state is presented in Fig.~\ref{fig:example_simos}c. The data demonstrates that the ground state follows a nearly linear trajectory with constant velocity. In contrast, the first excited state exhibits slight deviations from the linear trend. These deviations are attributed to the local expansion of the quantum dot at the points between tunnelling and plunger gates, allowing the electron in the excited state to redistribute its position.

This local increase in quantum dot size also results in a decrease in the energy splitting $(E_1 - E_0)$ at the same locations, as depicted in Fig.~\ref{fig:example_simos}d. This effect represents a potential weakness of the current optimisation algorithm: reduced energy separation increases the likelihood of non-adiabatic transitions to the excited state. Future improvements to the algorithm could involve incorporating additional optimisation strategies that maintain a constant $(E_1 - E_0)$ during the entire shuttling process to enhance the robustness of the transport.

\subsection{SiGe device}

To enhance sensitivity, charge sensors in Si/SiGe quantum dot devices \cite{van2024coherent, amitonov2025spinqubitperformanceerror} are operated in the multi-electron regime, typically containing approximately 400 electrons per sensor. However, the associated increase in charge density and the elevated bias applied to the sensor plunger electrodes induce significant distortion of the conduction band in the vicinity of the sensors. Consequently, realistic modelling of electron shuttling in Si/SiGe architectures (Fig.~\ref{fig:example_simos}e) must incorporate both sensors, substantially increasing the computational complexity compared to simplified models that neglect these effects, which were shown in Sections~\ref{sec:fdsoi}- \ref{sec:simos}.

The shuttling sequence was optimised using the algorithm outlined in Fig.~\ref{fig:algo}, to maintain the quantum dot energy at a constant value of -10meV while transporting the electron at a uniform velocity (Fig.~\ref{fig:example_simos}g,h). Throughout the shuttling process, the quantum dot's geometrical confinement remains largely unchanged. As a result, the energy level spacing remains relatively uniform, gradually decreasing from approximately 6~meV to 4~meV along the transport path. This behaviour arises from the high density of the dot array and the uniformity in gate geometry, which together ensure minimal variation in the electrostatic confinement potential.

\subsection{Computational Cost Analysis}

In this section, we summarise the computational cost for all models described above. All simulations were done on the AMD EPYC 7542 32-core CPU with 1Tb of RAM.

\begin{table}[ht]
\centering
\caption{\textbf{Computational cost of electron shuttling simulations.} 
Summary of runtime $\langle t \rangle$, number of nodes in the mesh $N$ and memory usage $M$ for different devices.}
\begin{tabular}{p{4cm}lccc}
\hline
Device & Classical charges & $\langle t\rangle$ (s) & $M$ (MB) & $N$\\
\hline
SiMOS device & Not included & 337.19 & 831.05 &723765\\
FD-SOI device without J-gates& In raised source/drain& 71.41 & 142.6 &152409  \\
FD-SOI device with J-gates&  In raised source/drain& 75.63 & 142.37 &152409 \\
Si/SiGe device& In sensors& 935.13 & 1903.73 &1719937 \\
\hline
\end{tabular}
\label{tab:performance}  

\end{table}

The summary of the computational cost analysis is in Tab.\,\ref{tab:performance}. It includes the average computation time of one iteration: loading of the mesh, solving the Poisson semiconductor equation, interpolating the conduction band edge to the Schr\"odinger solver, solving the Schr\"odinger equation and saving the results. The two FD-SOI devices have the lowest computational time, since the meshes are the smallest. The execution time of the SiMOS devices is also much lower than for Si/SiGe devices, not only because it has a lower number of nodes, but also because it doesn't include the simulation of classical charges.

\section{Conclusions}
We have developed and demonstrated an algorithmic approach for optimizing control voltages in conveyor-mode electron shuttling across various silicon-based quantum dot platforms. By ensuring a constant ground state energy and near-uniform transport velocity, the proposed method enhances the fidelity of quantum state transfer, which is essential for scalable quantum computing. Our simulations, applied to FD-SOI, SiMOS, and Si/SiGe device architectures, reveal that electrostatic confinement, gate overlap, and interface material properties significantly affect transport performance.

In FD-SOI devices, geometric limitations and interface-induced potential barriers lead to observable detuning and velocity oscillations, limiting the achievable transport fidelity. In contrast, SiMOS devices offer more uniform control but suffer from energy level variations that can increase non-adiabatic transition risk. Si/SiGe devices benefit from high dot density and regular gate geometry, enabling smoother transport, though modeling complexity increases due to the influence of nearby charge sensors.

Overall, our results demonstrate that accurate electrostatic modelling and waveform optimisation are crucial for coherent and efficient electron shuttling. Future work may extend this framework to include decoherence modelling, feedback-based optimisation, and real-time adaptive control for dynamic environments, pushing the boundaries of quantum dot interconnect technologies.


\bibliography{library}

\bibliographystyle{abbrv}

\end{document}